\documentclass[12pt]{article}
\usepackage{epsfig}
\usepackage{graphics}
\usepackage{psfrag}

\newcommand{\bea}{\begin{eqnarray}}
\newcommand{\eea}{\end{eqnarray}}
\newcommand{\simgt}{\hbox{ \raise3pt\hbox to 0pt{$>$}\raise-3pt\hbox{$\sim$} }}
\newcommand{\simlt}{\hbox{ \raise3pt\hbox to 0pt{$<$}\raise-3pt\hbox{$\sim$} }}

\newcommand{\LQ}{\Lambda_{\rm QCD}}

\begin{document}

\begin{titlepage}

    \begin{flushright}
      \normalsize TU-866, KEK-TH-1355\\
    April 12, 2010
    \end{flushright}

\vskip1.5cm
\begin{center}
\Large\bf\boldmath
Violation of Casimir Scaling for\\ 
Static QCD Potential 
at Three-loop Order
\unboldmath
\end{center}

\vspace*{0.8cm}
\begin{center}

{\sc C. Anzai}$^{a}$,
{\sc Y. Kiyo}$^{b}$ and
{\sc Y. Sumino$^{a}$}\\[5mm]
  {\small\it $^a$ Department of Physics, Tohoku University}\\[0.1cm]
  {\small\it Sendai, 980-8578 Japan}

  {\small\it $^b$ Theory Center KEK, Tsukuba}\\[0.1cm]
  {\small\it Ibaraki 305-0801, Japan}

\end{center}

\vspace*{0.8cm}
\begin{abstract}
\noindent
We compute the full ${\cal O}(\alpha_s^4)$ and
${\cal O}(\alpha_s^4\log\alpha_s)$ corrections to
the potential $V_R(r)$
between the static color sources, where
$V_R(r)$ is defined from the
Wilson loop
in a general representation $R$ of a general gauge group $G$.
We find a violation of the Casimir scaling of the
potential, for the first time, at ${\cal O}(\alpha_s^4)$.
The effect of the Casimir scaling violation is predicted to
reduce the tangent of $V_R(r)/C_R$ 
proportionally to specific color factors dependent on $R$.
We study the sizes of the Casimir scaling violation
for various $R$'s
in the case $G=SU(3)$.
We find that they are well within the present
bounds from lattice calculations,
in the distance region where both perturbative
and lattice computations of
$V_R(r)$ are valid.
We also discuss how to test the Casimir scaling violating effect.

\vspace*{0.8cm}
\noindent
PACS numbers: 12.38.Aw, 12.38.Bx, 14.40.Pq

\end{abstract}

\vfil
\end{titlepage}

\newpage

\section{Introduction}

The nature of the strong force is 
still a subject studied widely today.
In particular,
the static QCD potential  
has been studied extensively
for the purpose of elucidating the nature of the interaction between 
static color sources.
The static potential is a generalization of the Coulomb potential
in QED to the case of QCD.
Generally, the static 
potential at short-distances can be computed accurately
by perturbative QCD.
On the other hand,
the static potential at long-distances should be determined by
non-perturbative methods, such as
lattice simulations or 
computations based on various models.

Since some time, lattice computations have
shown that the static QCD potentials 
between the color sources in 
various color representations (in the color-singlet channel)
exhibit a property
known as ``Casimir scaling,'' within accuracies better than
5\%, and in the distance range 
$0.1~{\rm fm} \simlt r \simlt 1~{\rm fm}$
\cite{Markum:1988na,Bali:2000un}.\footnote{
See \cite{Ambjorn:1984mb,Liptak:2008gx}
for simulation studies on the Casimir scaling property for
gauge groups other than $SU(3)$.
}
Casimir scaling is a property of
the static potential $V_R(r)$
between the color sources in the representation $R$, that
the dependence of
$V_R(r)$ on $R$ is given only by an overall factor $C_R$,
the eigenvalue of the quadratic
Casimir operator for the representation $R$.
It turned out \cite{Shevchenko:2000du}
that the Casimir scaling property of the static potential
is a powerful discriminant of various models
and approaches, which attempt to explain the nature
of the QCD vacuum and color confinement.
So far, however, a reasoning of
Casimir scaling from the first principle has been
missing.

Computations of the
static potentials in perturbative QCD
have steadily made progress over decades.
In particular, the discovery of the cancellation of the 
renormalons in the total energy of a static quark-antiquark pair 
 led to a drastic improvement in
the accuracy of the perturbative prediction of the static 
potential between the fundamental charges \cite{Pineda:id}.
At the same time, the distance range, in which lattice 
computations and perturbative prediction agree with each other, extended
to a significantly wider range
\cite{Sumino:2001eh,Necco:2001xg,Pineda:2002se}.
By renormalization-group improvement of
the perturbative
prediction (after subtracting
the renormalon), this overlap range extended to
 $0.05~{\rm fm} \simlt r \simlt 0.4~{\rm fm}$ \cite{Sumino:2005cq}.
It was shown that 
non-perturbative contributions to the static potential 
in this overlap region are much smaller than previous
estimates, once the renormalon cancellation is incorporated
in the perturbative computations.
The linear potential (whose size is determined from the
long-distance behavior of the potential) is excluded as
a non-perturbative contribution in this distance region.
Furthermore, it was shown analytically that 
the perturbative static potential
approaches a ``Coulomb+linear'' form 
in the above distance range due to the higher-order terms
of the perturbative expansion, on the basis of a 
renormalon-dominance picture \cite{Sumino:2003yp}.

Recently two groups (including our group)
have independently completed computations of the 3-loop
corrections to the static potential between two fundamental
color representations \cite{Anzai:2009tm,Smirnov:2009fh}.
Compiling our present knowledge, the perturbative
expansion of the
static potential between the fundamental charges
is known up to ${\cal O}(\alpha_s^4)$ and also
all the logarithmic terms at ${\cal O}(\alpha_s^5)$
\cite{Pineda:2000gz,Brambilla:2006wp}
in this expansion are known.
Comparatively,
the perturbative expansion of the
static potential between the color sources in 
a general color representation
has been known only up to ${\cal O}(\alpha_s^3)$ \cite{Kauth:2009ud}.
The Casimir scaling is known to hold for the
perturbative potential at least up to this order.
 
In this paper we compute 
the full ${\cal O}(\alpha_s^4)$ and
${\cal O}(\alpha_s^4\log\alpha_s)$ corrections to the
static potential between the color sources in 
a general color representation, for a general
gauge group, by generalizing the computation for the
potential
for the fundamental representation.
The ${\cal O}(\alpha_s^4)$ correction originates
from the purely perturbative 3-loop correction, while
the ${\cal O}(\alpha_s^4\log\alpha_s)$ correction
originates from the ultra-soft correction.
We find that the Casimir scaling is violated, for the first
time, at ${\cal O}(\alpha_s^4)$.
In view of the consistency of
the perturbative and
lattice predictions for the fundamental potential,
it is interesting to
test consistency of the two predictions regarding
the Casimir scaling.
In particular, it is crucial whether the effects of
the violation of the Casimir scaling by perturbative QCD
exceed the current
bounds from lattice computations.
We examine the effects for various color representations
within perturbative QCD
and compare them with the lattice results.
We also discuss how to test the perturbative prediction
for the violation of the Casimir scaling.

In Sec.~2, after fixing our notations and  conventions,
we present our result for the ${\cal O}(\alpha_s^4)$ 
correction to the potential for a general color representation.
The ${\cal O}(\alpha_s^4\log\alpha_s)$ ultra-soft
correction, as well as the
complete expression for the
potential up to ${\cal O}(\alpha_s^4)$ 
and ${\cal O}(\alpha_s^4\log\alpha_s)$,
are given in Sec.~3.
Sec.~4 gives an analysis of the color factors in
perturbative computations
up to 3 loops.
We examine the effects of the Casimir scaling violation 
for various representations in Sec.~5.
Concluding remarks are given in Sec.~6.

\section{Wilson loop and 3-loop static potential}

We consider the (vector)
gauge theory for a general gauge group $G$
with $n_l$ massless fermions in the fundamental
representation.
In the case $G=SU(3)$, the theory is QCD with $n_l$
flavors of massless quarks, which is our main concern.
Nevertheless, to keep generality, all our formulas will be
presented for the general gauge group.

\begin{figure}[t]
  \begin{center}
  \vspace*{-0.2cm}
  \includegraphics[width=0.4\textwidth]{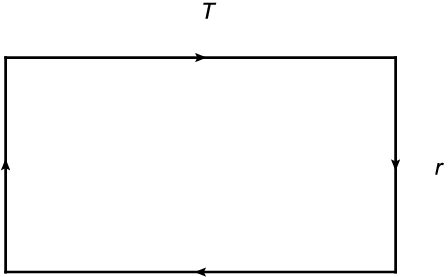}
  \vspace*{-0.2cm}
  \caption{Contour $C$ for the Wilson loop in $W_R[C]$.}
  \label{fig1}
  \end{center}
\end{figure}
We define the static potential between the color sources
in a general representation $R$, in the singlet channel.\footnote{
Potentials in non-singlet channels are defined by
inserting appropriate projection operators on the Wilson loop at
temporal boundaries.
Presently the potential for the fundamental representation
in the octet channel is known up to 2-loop order \cite{Kniehl:2004rk}.
}
For this purpose we first define the vacuum expectation value 
of the Wilson loop as
\begin{eqnarray}
W_R[C]&=&
\langle 0| \,{\rm \hat{\cal T}}\biggl[
{\rm Tr}\,{\rm P}
\exp
\left\{ig_s \oint_{C} dz^\mu A_\mu^a(z)T_R^a\right\}
\biggr]
|0\rangle/
\langle 0|{\rm Tr}\,{\bf 1}_{R}|0\rangle
\,.
\label{WL}
\end{eqnarray}
Here,
${\rm \hat{\cal T}}$ denotes the time-ordering of the gauge field
operators $ A_\mu^a(z)$.
The hermitian generators of a general representation $R$ are
denoted by $T_R^a$, which satisfy the commutation relation
$[T^a_R, T^b_R]= if^{abc} T_R^c$ with
the structure constant $f^{abc}$ of the gauge group $G$.
The fundamental and adjoint representations, 
respectively, will be denoted explicitly
by $R=F$ and $R=A$.
The trace ${\rm Tr}$ is taken
in the representation $R$, and ${\bf 1}_R$ denotes
the identity. 
${\rm P}$ stands for the path-ordering of $T_R^a$
along the contour $C$, which 
is a rectangular loop of spatial extent $r$ and
time extent $T$; see Fig.~\ref{fig1}.

\begin{table}
\begin{tabular}{c|cc}
\hline \hline
 & $SU(N)$ & $SO(N)$ \\
\hline
$N_F$  & $N$ & $N$ \\
$N_A$  & $N^2-1$ & $N(N-1)/2$ \\
$C_F$  & $(N^2-1)/(2N)$ & $(N-1)/4$ \\
$C_A$  & $N$ & $(N-2)/2$ \\
$T_A$  & $N$ & $(N-2)/2$ \\
$d_{F}^{abcd}d_F^{abcd}/(N_A T_F)$   &
$(N^4-6N^2+18)/(48 N^2)$ & 
~~~$(N^2-N+4)/192$~~~ \\
$d_{F}^{abcd}d_A^{abcd}/(N_A T_F)$   &
$N(N^2+6)/24$  & 
$(N-2)(N^2-7N+22)/192$ \\
$d_{F}^{abcd}d_A^{abcd}/(N_A T_A)$   &
$(N^2+6)/48$  &  $(N^2-7N+22)/192$ \\
~~~$d_{A}^{abcd}d_A^{abcd}/(N_A T_A)$~~~  &
$N\,(N^2+36)/24$ & 
$(N^3-15N^2+138N-296)/192$ \\
\hline \hline
\end{tabular}
\caption{\small
Some of the color factors related to the
fundamental (vector) and adjoint representations
 ($R=F$ and $R=A$, respectively) in the case 
$G=SU(N)$ and $G=SO(N)$.
Our convention is $T_F=1/2$.
}
\label{tab:color}
\end{table}

The normalization of $T^a_R$ is given by
\bea
{\rm Tr}(T^a_R T^b_R)= T_{R}\, \delta^{ab} .
\eea
We fix our convention by setting $T_F=1/2$.
The dimension of the 
representation $R$ is denoted by $N_R$.
The quadratic Casimir operator as well as symmetric
invariant tensors which will appear in our computation
are defined as 
\begin{eqnarray}
&&
(T_R^a T_R^a)_{ij}=C_R \delta_{ij},
\label{CasimirOp}
\\ &&
d_R^{a_1\dots a_n}=
\frac{1}{n!}\sum_\pi
{\rm Tr}
\left(
T_R^{a_{\pi(1)}}
\dots
T_R^{a_{\pi(n)}}
\right),
\label{invtensor}
\end{eqnarray}
where the sum 
is over all permutations $\pi$ of the indices.
There are a number of relations satisfied by these
color factors; see \cite{vanRitbergen:1998pn}.
We note two of them: (i) $T_R N_A=C_R N_R$, which follows
from eq.~(\ref{CasimirOp}) after taking trace;
(ii) $d_A^{a_1\dots a_n}=0$ for odd $n$'s, since $(T^a_A)^T=-T^a_A$.
Some values of the color factors are listed in Tab.~\ref{tab:color}
for $G=SU(N)$ and $G=SO(N)$.

The static potential between the color sources
in a general representation $R$
is defined from the Wilson-loop expectation value as
\begin{eqnarray}
V_{R}(r)
&=&
\lim_{T\rightarrow \infty}\,\frac{1}{(-iT)}\,
\log(W_R[C]) \,.
\label{eq:def-VR}
\end{eqnarray}
In other words, the potential appears in $W_R[C]$ in an exponentiated form
for large $T$, $W_R[C]\sim \exp\{-iTV_R(r)\}$.

We may evaluate
$V_{R}(r)$ using perturbation theory.
Since our technology for loop computations is developed mostly in
momentum space, we compute the potential in momentum space.
In order to regularize both
UV and IR divergences, we employ dimensional regularization with one temporal dimension
and $d = D-1 = 3-2\epsilon$ spatial dimensions.
Thus, the perturbative potential expressed in terms
of the corresponding ($V$-scheme)
coupling in momentum space reads
\bea
{V}_R^{\rm PT}(r)
&=&
\left(\frac{\mu^2e^{\gamma_E}}{4\pi}\right)^\epsilon
\int \frac{d^d\vec{q}}{(2\pi)^d}\,
e^{i\vec{q}\cdot \vec{r}}\,
\left[
-4\pi C_R \frac{\alpha_{V_R}^{\rm PT}(q)}{q^2}\,
\right]\,,
\label{eq-VR-PT}
\eea
where both quantities are denoted with superscripts PT
to make explicit that
they are computed in perturbative expansions of the strong coupling
constant.
A prefactor is included on the right-hand side
such that $\alpha_{V_R}^{\rm PT}(q)$ is
defined to be dimensionless;
$q=|\vec{q}|$;
$\gamma_E = 0.5772...$ denotes the Euler constant.

The perturbative expansion of $\alpha_{V_R}^{\rm PT}(q)$
is expressed as
\bea
&&
\alpha_{V_R}^{\rm PT}(q)
= \alpha_s(\mu) \, \sum_{n=0}^{\infty} P_n(L ) \,
\biggl( \frac{\alpha_s(\mu)}{4\pi} \biggr)^n
\label{alfVPT}
\eea
with
\bea
&&~~~
L=\log \Biggl(\frac{\mu^2}{q^2}\Biggr).
\eea
Here, $\alpha_s(\mu)$ denotes the strong coupling constant
renormalized at the renormalization scale $\mu$,
defined in the modified minimal subtraction ($\overline{\rm MS}$) scheme;
$P_n(L)$ denotes an $n$th-degree polynomial of $L$.
The renormalization-group equation of $\alpha_s(\mu)$ is
given by
\begin{eqnarray}
\frac{d}{d\log(\mu^2)}\left(\frac{\alpha_s(\mu)}{4\pi}\right)
&=&-\sum_{n=-1}^\infty \beta_n\,
\left(\frac{\alpha_s(\mu)}{4\pi}\right)^{n+2}\,,
\end{eqnarray}
where $\beta_n$ represents the $(n+1)$-loop coefficient of the
beta function.
The relevant coefficients read \cite{Tarasov:1980au}
\begin{eqnarray}
&&\beta_{-1}=\epsilon\,, ~~~
\beta_0=\frac{11}{3}\,C_A-\frac{4}{3}\,n_l\,T_F\,,
~~~
\beta_1=\frac{34}{3}\,C_A^{\,2}
-\left(\frac{20}{3}\,C_A+ 4\,C_F \right)\,n_l T_F\,,
\nonumber\\
&&
\beta_2=
\frac{2857}{54}\,C_A^{\,3}
-\left( \frac{1415}{27}\,C_A^2
       +\frac{205}{9}\, C_A C_F
       -2\,C_F^2
\right)\, n_l T_F
\nonumber\\
&& ~~~~~~~
+\left(\frac{158}{27}\,C_A+\frac{44}{9}\,C_F\right)\, n_l^2 T_F^{\,2} 
\,.
\end{eqnarray}
Both ${V}_R^{\rm PT}(r)$ and
$\alpha_{V_R}^{\rm PT}(q)$ are independent of the scale $\mu$,
hence their dependences on $\log \mu^2$
are dictated by the renormalization-group equation.
For $n\leq 2$, the only part of the polynomial $P_n(L )$
that is not determined by the renormalization-group equation is $a_n \equiv P_n(0)$.
For $n\geq 3$, $P_n(L)$ includes IR divergences in terms
of poles of $\epsilon$ and assoicated logarithms, whose coefficients are
not determined by $\beta_i$'s.
Up to 3--loop order, they are given by
\bea
&&
P_0 = a_0 ,~~~
P_1 = a_1 + a_0\beta_0  L ,
~~~
P_2 = a_2 + (2a_1\beta_0 + a_0\beta_1 )L +
a_0{\beta_0}^2 L^2 ,
\nonumber \\ &&
{P}_3 =
a_3  + ( 3a_2\beta_0+
   2 a_1 \beta_1 +
    a_0 \beta_2) L
+
   \biggl(3 a_1 {\beta_0}^2
+ \frac{5}{2}  a_0 \beta_0 \beta_1\biggr) L^2
+  a_0 {\beta_0}^3  L^3 \,,
\eea
and
\bea
&&
a_3=\bar{a}_3+\frac{8}{3}\pi^2C_A^3\biggl(
\frac{1}{\epsilon}+3\,L
\biggr) .
\label{a3}
\eea
The $1/\epsilon$ term represents
the IR divergence \cite{Appelquist:es,Brambilla:1999qa,Kniehl:2002br}, 
which is separated
together with the corresponding scale dependence.

The first three $a_n$'s are independent of the
representation $R$ \cite{Kauth:2009ud}:
\begin{eqnarray}
&&
a_0=1\,,
~~~
a_1 =
\frac{31}{9}\,C_A -\frac{20}{9}\,T_F\,n_l\,,
\nonumber\\&&
a_2 =
 \left(\frac{4343}{162}+4\pi^2-\frac{\pi^4}{4}+\frac{22}{3}\zeta_3\right)\,C_A^{\,2}
-\left( \frac{1798}{81}+\frac{56}{3}\zeta_3\right)\,C_A T_F n_l
\nonumber \\ &&
~~~~~~
-\left(\frac{55}{3}-16\zeta_3 \right)\,C_F T_F n_l
+\left(\frac{20}{9}T_F n_l\right)^2\,,
\label{eq:a1-a2}
\end{eqnarray}
where $\zeta_3=\zeta(3)=1.2020...$ denotes the Riemann zeta function 
$\zeta(z)=\sum_{n=1}^\infty {1}/{n^z}$ evaluated
at $z=3$.
Since $C_R$ is factored out in eq.~(\ref{eq-VR-PT}), 
and since $\beta_n$'s are independent of the representation $R$, if
all $a_n$'s are also independent of $R$,
the potential ${V}_R^{\rm PT}(r)$ satisfies the Casimir scaling.

The 3-loop non-logarithmic constant
$\bar{a}_3$ for the fundamental color
 sources has been computed recently:
the contributions with internal fermion loops in \cite{Smirnov:2008pn} and
the purely gluonic contributions in \cite{Anzai:2009tm,Smirnov:2009fh}. 
We repeat these calculations
for the general color sources.
We find
\begin{eqnarray}
\bar{a}_3
&=&
-\left(\frac{20}{9}n_l T_F\right)^3 \!
+\bigg[
       C_A\left( \frac{12541}{243}
                +\frac{64\pi^4}{135}
                +\frac{368}{3}\zeta_3\right)
   +C_F\left( \frac{14002}{81}
                -\frac{416}{3}\zeta_3\right)
 \bigg]\,(n_l T_F)^2
\nonumber \\
&+&
\bigg[\,
  2\,c_1\, C_A^{\,2}
 +\left(-\frac{71281}{162}+264\zeta_3+80\zeta_5\right)C_A\, C_F
 +
\left(\frac{286}{9}+\frac{296}{3}\zeta_3-160\zeta_5\right) C_F^{\,2}
\bigg]\,n_l\,T_F
\nonumber \\
&+&
\frac{1}{2}\,c_2\,n_l\, \biggl(\frac{d_F^{abcd}d_R^{abcd}}{N_A\,T_R}\biggr)\,
+\bigg[\,
c_3\, C_A^{\,3}
+\frac{1}{2}\,c_4\, \biggl(\frac{d_A^{abcd}d_R^{abcd}}{N_A\,T_R}\,\biggr)
\bigg]
\,,
\end{eqnarray}
where $\zeta_5=\zeta(5)=1.0369...$.
In the above equation, the coefficients
$c_i$'s are known only numerically:
\bea
&&
c_1 = -354.859, ~~~ c_2=-56.83(1), ~~\cite{Smirnov:2008pn}
\eea
and\footnote{
The results of Ref.~\cite{Smirnov:2009fh} are
$c_3=502.24(1)$, $c_4=-136.39(12)$.
}
\bea
&&
c_3=502.22(12) 
\,,
~~~
c_4=-136.8(14) ~~\cite{Anzai:2009tm}
\,.
\eea
We find that $\bar{a}_3$ is dependent
on the representation $R$
through the color factors $d_F^{abcd}d_R^{abcd}/(N_AT_R)$
and  $d_A^{abcd}d_R^{abcd}/(N_AT_R)$.
These are the leading effects of the Casimir scaling violation
in perturbation theory.

One may want to include massless
fermions in representations other than
the fundamental representation.
One may easily identify the
contributions from fermions in
a general representation $R'$ by simple replacements
$T_F\to T_{R'}$, $C_F\to C_{R'}$ and $d_F^{abcd}\to d_{R'}^{abcd}$
in the coefficients of $n_l$ in the
equations presented in this section.

\section{Ultra-soft correction}

The IR divergence contained in $a_3$
is an artifact of the strict perturbative expansion
of the static potential ${V}_R(r)$   in $\alpha_s$.
The divergence originates from the degeneracy of
the singlet and adjoint intermediate states 
in $W_R[C]$ within the naive perturbation theory \cite{Appelquist:es}.
Beyond (naive) perturbation theory, this degeneracy is known to be lifted
and the IR divergence is absent.
The difference between the static potential
${V}_R(r)$ and its perturbative expansion
${V}_R^{\rm PT}(r)$ can be treated systematically
within the effective field theory
Potential Non-relativistic
QCD (pNRQCD)\footnote{
Strictly speaking,
the name of the effective theory, as well as terminology such as
``quarks'' and ``gluons'', are adequate only in the case $G=SU(3)$.
We will frequently use them for the general gauge group $G$, nonetheless.
} 
\cite{Brambilla:2004jw},
which treats ultra-soft (US) gluons and heavy quark-antiquark 
bound-states as dynamical degrees of freedom.
This difference ${V}_R(r)-{V}_R^{\rm PT}(r)$
is given by contributions of ultra-soft degrees
of freedom.

We repeat the computation of the ultra-soft contributions
given in \cite{Brambilla:1999qa,Kniehl:1999ud} for the case
of ${V}_R(r)$, taking into account the following points:
(1) We change the representation of the static charge of
the Wilson loop from the fundamental to the general representation $R$.
(2) The perturbative matching between the full theory and pNRQCD
is properly taken into account, such that the non-logarithmic term
of ${V}_R(r)-{V}_R^{\rm PT}(r)$ is predicted correctly 
at ${\cal O}(\alpha_s^4)$.
Thus, we have
\begin{eqnarray}
\delta V_R^{\rm US}(r)
&\equiv&
{V}_R(r)-{V}_R^{\rm PT}(r)
\nonumber\\
&=&
-i g_s^2\, \frac{T_R}{N_R}
\int_0^\infty dt \, e^{-it\,(V_{\rm ad}-V_s)}
\nonumber\\
&&\times
\langle 0| \,{\rm \hat{\cal T}}\biggl[
{\bf r}\cdot{\bf E}^a(t)
\left( {\rm P}\,
e^{ig_s\int_{0}^t dt'\, A_0^c(t') T_A^c}\right)_{ab}\,
{\bf r}\cdot{\bf E}^b(0)
\biggr]
|0\rangle
+{\cal O}(r^3)
\,.
\label{eq:VR-US}
\end{eqnarray}
The term shown explicitly is the leading-order 
[${\cal O}(r^2)$] term in the multipole
expansion in ${\bf r}$, where $r=|{\bf r}|$. 
All fields in the matrix
element are evaluated at the spatial origin.
$(T_A^c)_{ab}=-if^{abc}$ denotes the generator of the
adjoint representation.
$V_s$ and $V_{\rm ad}$ denote the singlet and adjoint potentials, respectively,
in the representation $R$, i.e., 
the potentials between the sources in the singlet and adjoint
channels, respectively, in $R\otimes \bar{R}$.

In the distance region $r\ll \LQ^{-1}$, 
we may further expand the matrix element as well as
$V_{\rm ad}-V_s$
in the coupling constant and obtain
the leading--order contribution to
$\delta V_R^{\rm US}(r)$ in 
double expansion in $\alpha_s$ and $\log(\alpha_s)$.\footnote{
One can verify that indeed $\delta V_R^{\rm US}(r)
={V}_R(r)-{V}_R^{\rm PT}(r)$
vanishes in the strict perturbative expansion, by expanding also
$e^{-it\,(V_{\rm ad}-V_s)}$ in $\alpha_s$;
in this case, each integral becomes scaleless, and
therefore it vanishes in dimensional regularization.
}$^,$\footnote{
${\cal O}(r^3)$ terms in eq.~(\ref{eq:VR-US}) do not
contribute to the leading order of this double expansion.
}
It is understood that we employ dimensional regularization,
and $g_s$ can be identified with the bare coupling constant 
of the full theory in the leading order.
Noting that the
singlet and adjoint potentials are given by
\begin{eqnarray}
V_s(r)
&=&
-\frac{C_R\,\alpha_s(\mu)}{r}\,
\left[
\frac{\left(\mu^2\, r^2\, e^{\gamma_E}\right)^\epsilon\,
\Gamma(1-2\epsilon)}{\Gamma(1-\epsilon)}
\right]
+{\cal O}(\alpha_s^2)\,,
\label{eq:Vs-tree}
\\
V_{\rm ad}(r)
&=&
-\frac{\left(C_R-\frac{1}{2}\,C_A\right)\,\alpha_s(\mu)}{r}\,
\left[
\frac{\left(\mu^2\, r^2\, e^{\gamma_E}\right)^\epsilon\,
\Gamma(1-2\epsilon)}{\Gamma(1-\epsilon)}
\right]
+{\cal O}(\alpha_s^2)\,,
\label{eq:Vo-tree}
\end{eqnarray}
we obtain
\begin{eqnarray}
\Bigl[ \delta V_R^{\rm US}(r)\Bigr]_{\rm LO}
&=&
-g_s^2\,C_R\frac{d-1}{d}\,
r^2\,
\left(\frac{(V_{\rm ad}-V_s)^2}{4\pi}\right)^{{d}/{2}}\,
\left[\frac{\Gamma(1+d)\Gamma(-d)}{\Gamma(\frac{d}{2})}\right]\,
\nonumber\\
&=&
\frac{C_R\,C_A^{\,3}\, \alpha_s(\mu)^4}{24\pi r}
\left[ \,
  \frac{1}{\epsilon}
  +4\log \left(\mu^2r^2 \right)
  -2\log\left(C_A\alpha_s(\mu)\right)
  +\frac{5}{3}+6\gamma_E \,
\right]\,,
\label{deltaEUS-pert}
\end{eqnarray}
where we dropped ${\cal O}(\epsilon)$ terms in the last expression.
Since $[ \delta V_R^{\rm US}]_{\rm LO}$ 
depends on the representation $R$ only through
the overall factor $C_R$, Casimir scaling is preserved by
this contribution.
The $1/\epsilon$ term represents the UV divergence of
$[ \delta V_R^{\rm US}]_{\rm LO}$ within pNRQCD effective
theory.
Upon Fourier transform, the $1/\epsilon$ and $\log (\mu^2)$ terms
of eqs.~(\ref{a3}) and (\ref{deltaEUS-pert})
cancel each other.

\begin{figure}[t]\centering
\includegraphics[width=8cm]{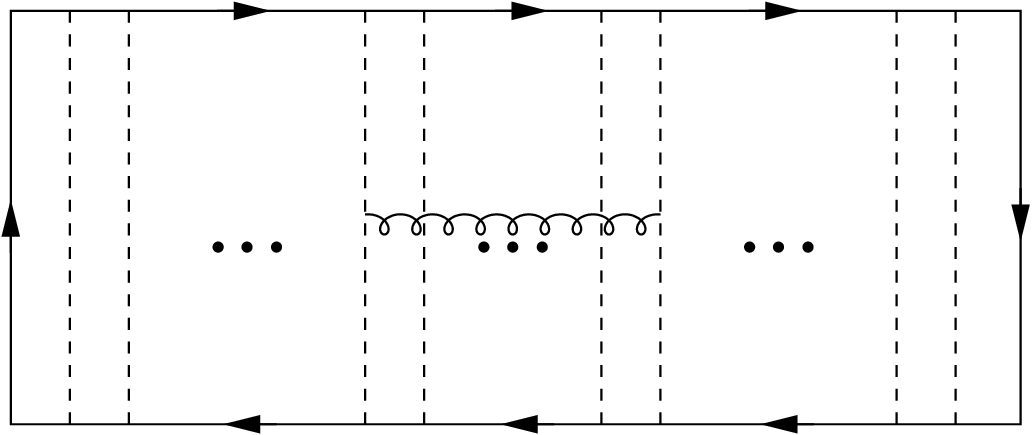}
\caption{\small 
Class of diagrams contributing to $[ \delta V_R^{\rm US}]_{\rm LO}$.
Dashed lines represent Coulomb gluons, while
a curly line represents transverse gluon, in Coulomb gauge.
\label{Coulomb-resum}}
\end{figure}
We made a cross check of eq.~(\ref{deltaEUS-pert})
as follows.
It was first pointed out in \cite{Appelquist:es} 
that $[ \delta V_R^{\rm US}]_{\rm LO}$ 
may be obtained 
as the difference between the
resummation of Coulomb ladder diagrams,
given in Fig.~\ref{Coulomb-resum}, and the sum of the
individual diagrams (in Coulomb gauge).\footnote{
To be precise, one should replace $C_A/2-C_R$ by $C_A/2$
in the latter (individual) diagrams,
in order to take into account the shift of the ground-state
energy $0\to V_s(r)$ by Coulomb resummation.
}
Namely, $[ \delta V_R^{\rm US}]_{\rm LO}$ may be computed
more directly using a resummation of diagrams
in perturbation theory, without recourse
to pNRQCD effective theory.
One can verify that this is indeed the case, including the
non-logarithmic terms in eq.~(\ref{deltaEUS-pert}).
Formally the equivalence of the two approaches 
can be shown using the threshold expansion
of Feynman diagrams \cite{Beneke:1997zp}.
We have also evaluated the relevant diagrams
directly and confirmed the coefficients of the logarithms
as well as the
non-logarithmic constant
explicitly.
In the resummation approach, one can avoid complexity
related to the matching of the
full theory to pNRQCD,
in computing the non-logarithmic terms.
(See also \cite{Beneke:2007pj}.) 

For completeness,
we present the formula for ${V}_R(r)$,
as given by
the sum of ${V}_R^{\rm PT}(r)$ and
$\delta V_R^{\rm US}(r)$,
including all the corrections up to ${\cal O}(\alpha_s^4)$ and
${\cal O}(\alpha_s^4\log\alpha_s)$.
Namely we use the series expansion (\ref{alfVPT})
up to $n=3$ for the former and
eq.~(\ref{deltaEUS-pert}) for the latter.
We obtain
\bea
&&
\Bigl[{V}_R(r)\Bigr]_{\rm NNNLO}
=-C_R\frac{\alpha_s(\mu)}{r}
\, \sum_{n=0}^{3} \widetilde{P}_n({L}' ) \,
\biggl( \frac{\alpha_s(\mu)}{4\pi} \biggr)^n \,,
\eea
where
\bea
{L}'=\log(\mu^2 r^2) + 2{\gamma_E}\,,
\eea
and
\bea
&&
\widetilde{P}_0 = a_0 ,~~~
\widetilde{P}_1 = a_1 + a_0\beta_0  L' ,
~~~
\widetilde{P}_2 = a_2 + (2a_1\beta_0 + a_0\beta_1 )L' +
a_0{\beta_0}^2 \biggl(L'^{\,2}+\frac{\pi^2}{3}\biggr) ,
\nonumber \\ &&
\widetilde{P}_3 =
\bar{a}_3 +\delta a_3^{\rm US} + 
( 3a_2\beta_0+   2 a_1 \beta_1 +
    a_0 \beta_2) L'
\nonumber\\ &&
~~~~~~~~
+
   \biggl(3 a_1 {\beta_0}^2
+ \frac{5}{2}  a_0 \beta_0 \beta_1\biggr) \biggl(L'^{\,2}+\frac{\pi^2}{3}\biggr)
+  a_0 {\beta_0}^3 (L'^{\,3}+\pi^2 L'+16\zeta_3)
\,,
\eea
with
\bea
&&
\delta a_3^{\rm US} = \frac{16}{3}\pi^2C_A^3
\biggl[ \log \Bigl(C_A\alpha_s(\mu)\Bigr) + \gamma_E -\frac{5}{6}
\biggr] \,.
\eea
By definition $[{V}_R(r)]_{\rm NNNLO}$ is a renormalization-group invariant quantity and is
free from IR divergences.
Apart from the overall factor $C_R$, only $\bar{a}_3$ in the
above expression is dependent on
the representation $R$, which is the source of the violation
of the Casimir scaling.

Let us comment on the non-perturbative contributions to
the static potential
$V_R(r)$.
At $r< \LQ^{-1}$, the multipole expansion in $r$ of
$\delta V_R^{\rm US}(r)$,
given in eq.~(\ref{eq:VR-US}), constitutes
an operator-product-expansion of
$\delta V_R^{\rm US}(r)$. 
In this way, one can systematically parametrize the 
non-perturbative contributions to $\delta V_R^{\rm US}(r)$
[and therefore to $V_R(r)$]
in terms of matrix elements of non-local gluon condensates, after
subtracting UV divergences.
See \cite{Brambilla:1999qa,Pineda:2002se,Sumino:2005cq,Koma:2006si} 
for analyses of the non-perturbative contributions
to $V_R(r)$ in this framework.
$[ \delta V_R^{\rm US}]_{\rm LO}$ can be regarded as 
defining
a scheme for subtracting the UV divergence at leading order,
or equivalently, as defining a scheme for the corresponding
non-perturbative contributions.

\section{Color factors in loop diagrams}

We examine the general structure, especially
the $R$ dependence, of the color factors
in ${V}_R^{\rm PT}(r)$ without going into details of
the loop integrals.
A discussion on this subject up to 2 loops is given 
briefly in
\cite{Kauth:2009ud}.
We extend the argument to 3 loops and to more details.

\begin{figure}[t]\centering
\includegraphics[width=2.5cm]{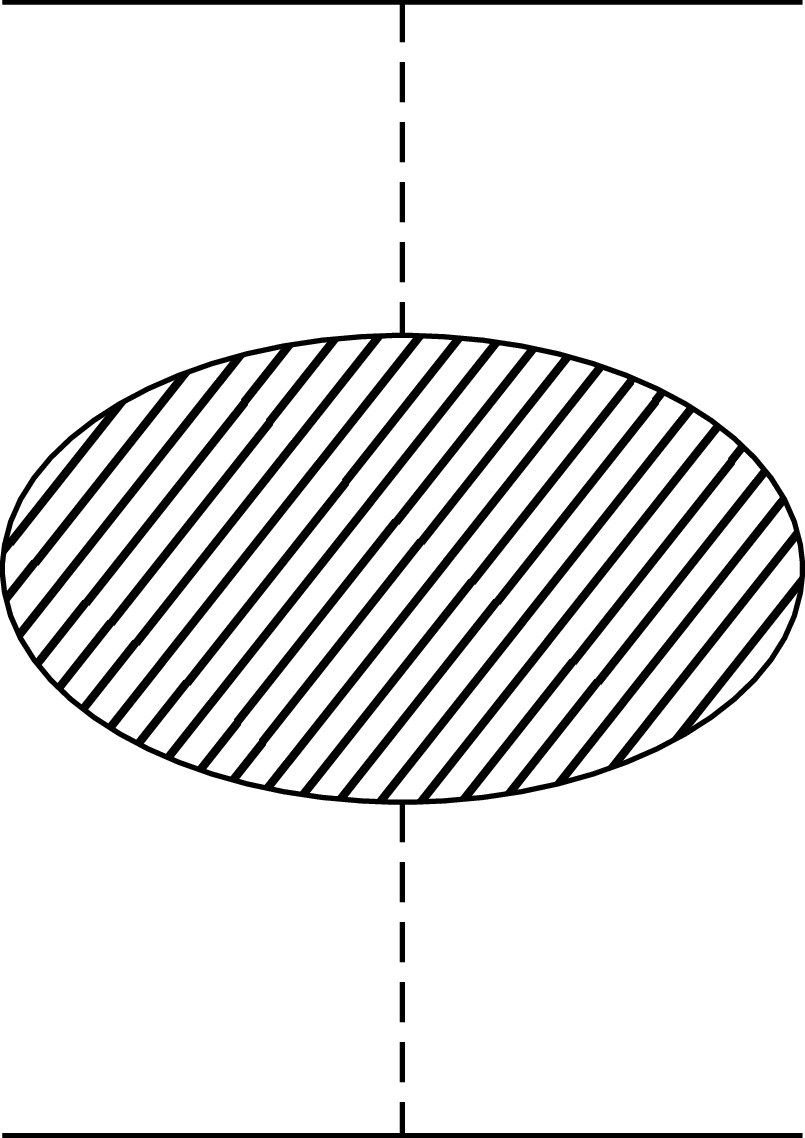}
\caption{\small 
Class of diagrams contributing to the potential
at 1-loop in Coulomb gauge.
The hatched part represents the vacuum polarization of
the Coulomb gluon.
\label{Coulomb-vac-pol}}
\end{figure}

At tree-level, the color factor of the potential is simply
${\rm Tr}(T^a_RT^a_R)/{\rm Tr}({\bf 1}_R)=C_R$.
As is well-known, 
a simple way to understand the color factor at 1-loop order is
to consider the corrections in Coulomb gauge.
In this gauge,
only the vacuum polarization of the Coulomb gluon 
contributes to the potential \cite{Feinberg:1977kc}; see Fig.~\ref{Coulomb-vac-pol}.
Since the vacuum polarization is independent of $R$ and
is proportional to $\delta^{ab}$,
the $R$-dependence of the potential is the same as the
tree-level potential.
The same argument cannot be used for the 2-loop or
higher-order corrections,
however, since there are contributions from graphs other than the
vacuum-polarization type.
Hence, we develop a more general argument.

Let us make some preparations.
First, according to eq.~(\ref{invtensor}),
${\rm Tr}(T^{a_1}_R\cdots T^{a_n}_R)$
equals $d_R^{a_1\dots a_n}$ plus a sum of the terms
which include at least one commutator
$[T^{a_i}_R,T^{a_j}_R]$.
If we rewrite the commutator as $if^{a_ia_jb}T_R^b$ and
repeat this operation recursively,
we may express the trace by $d_R^{a_1\dots a_n}$ and
lower rank tensors as
\bea
{\rm Tr}(T^{a_1}_R\cdots T^{a_n}_R) =
d_R^{a_1\dots a_n} + \sum_{k=2}^{n-1} c\cdot
d_R(k) \,,
\label{prep1}
\eea
where 
in the second term on the right-hand side we suppressed
the color indices;
$d_R(k)$ represents a $k$th-rank invariant tensor;
$c$ is an $R$-independent coefficient and expressed in terms of $f^{abc}$.
($d_R(k)$ and $c$ have appropriate color indices which are not shown.)
Contracting both sides with $f^{a_ia_jb}$, where
$\{a_i,a_j\}\subset\{a_1,\dots,a_n\}$, the first term on the right-hand
side drops. 
Hence, we have
\bea
f^{a_ia_jb}\,{\rm Tr}(T^{a_1}_R\cdots T^{a_n}_R) =
\sum_{k=2}^{n-1}\sum c'\cdot
d_R(k) \,,
\label{prep2}
\eea
where we absorbed $f^{a_ia_jb}$ in $c'$.
Note that on the left-hand side there are $n$ generators $T^a_R$,
while  the
highest rank of the
invariant tensor on  the right-hand side is $n-1$.


In perturbative computations of the static potential,
one first eliminates iterations of lower-order potentials
from the diagrams contributing to the expectation value of the Wilson loop,
$W_R[C]$.
A prescription has been developed, in association with a proof
for the exponentiation of the static potential \cite{Gatheral:1983cz}.
Let us briefly review the prescription.
Within the diagrams for
$W_R[C]$, only the 2-particle-irreducible (2PI) diagrams contribute
to the potential.
Here, 2-particle irreducibility is defined with respect to cutting the
two static time-like lines simultaneously.
Furthermore, among the 2PI diagrams,
the color factors of the diagrams, which become 2-particle reducible
(2PR) by sliding vertices on the two
static lines, need to be modified;
this is because, part of these diagrams are identified with
iterations of lower-order potential.\footnote{
If the gauge group is abelian, any
diagram, which becomes 2PR
by sliding vertices on the two time-like lines, 
does not contribute to the potential.
This is because, these diagrams (at a fixed order)
sum up exactly to the exponentiation
of the lower-order diagrams which contribute to the potential, 
if we can ignore orderings
of $T^a_R$'s.
}
How to modify the color factors of these diagrams
is the essence of the
prescription.
One associates a ``color graph'' with each of these diagrams.
Let us denote the original 2PI diagram by $D$ and the corresponding
color graph by $C_0(D)$.
To begin with, the topology of the color graph $C_0(D)$
is taken to be the same as that of $D$, and the value 
of $C_0(D)$ is set equal to
the color factor of the diagram $D$.
Then one tries to render the color graph $C_0(D)$ to a 2PR graph
by sliding vertices on the two static lines.
\begin{figure}[t]\centering
\includegraphics[width=15cm]{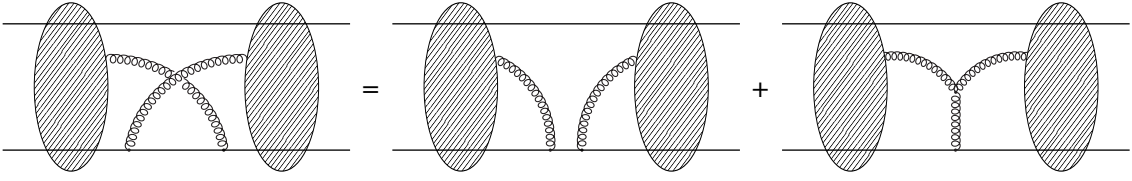}
\caption{\small 
Reduction of a 2PI color graph to a 2PR graph and a
residual graph using
the identity
$T^a_RT^b_R=T^b_RT^a_R+if^{abc}T^c_R$.
The 2PR graph (first graph on the right-hand side) will be deleted.
In the second graph, the number of vertices on the static
line is reduced by one.
\label{colorgrf-reduct}}
\end{figure}
Whenever any two vertices $a$ and $b$ on a same line
are interchanged in this procedure, one separates $C_0(D)$ into
two color graphs $C_1(D)$ and $C_2(D)$, corresponding to 
$T^a_RT^b_R=T^b_RT^a_R+[T^a_R,T^b_R]$.
In the latter graph, the two vertices are combined to a
single vertex on the line, corresponding to $if^{abc}T^c_R$;
see Fig.~\ref{colorgrf-reduct}.
Thus, the value of $C_2(D)$ equals the value of $C_0(D)$
except that
$T^a_RT^b_R$ is replaced by $if^{abc}T^c_R$.
This procedure is repeated for each color graph, 
until no more reduction of any of the 2PI graphs 
into a 2PR graph is possible.
Then one deletes all the 2PR graphs.
The sum of the values of the remaining 2PI 
color graphs $\sum_{i \in {\rm 2PI}} C_i(D)$ is the color factor
that should be assigned to the original diagram $D$.

After applying
the above prescription, (the value of) any of the color graphs $C_i(D)$,
which correspond to a general 2PI diagram $D$, can be written 
in the form
\bea
&&
C_i(D)={\rm Tr}({T^{a_1}_R\cdots T^{a_n}_R})\,
X_n^{a_1\dots a_n}
\nonumber\\
&&
~~~~~~~~
=N_RC_R\times\left[
\frac{X_2^{aa}}{N_A}+
\frac{1}{N_AT_{R}}
(d_R^{abc}X_3^{abc}+\cdots +
d_R^{a_1\dots a_n}X_n^{a_1\dots a_n})
\right]
\,,
\label{geneformofC}
\eea
where $X_k^{a_1\dots a_k}$'s are independent of $R$.
In the second equality,
we substituted eq.~(\ref{prep1}) and used
$d_R^{ab}=T_{R}\delta^{ab}$,
$T_R=N_RC_R/N_A$.
The overall factor $N_R$ will be canceled in the potential
by the denominator ${\rm Tr}({\bf 1}_R)=N_R$.
Each term of $X_k^{a_1\dots a_k}$ is not factorizable
(e.g.\ $X_4^{abcd}$ does not include a term of the form $A^{ab}B^{cd}$),
since a factorizable term corresponds to a 2PR color graph.\footnote{
2PR color graphs, which are deleted, 
contain powers
of $C_R$ if they contain (for instance) iterations of the
tree graph, 
$X_k^{a_1\dots a_k}\propto \delta^{a_1a_2}\delta^{a_3a_4}\cdots$,
since $T_R^aT_R^a=C_R{\bf 1}_R$.
}
We will see this in explicit examples below.

We note that, in eq.~(\ref{geneformofC}),
purely gluonic contributions to $X_k^{a_1\dots a_k}$ 
(expressed in terms of only $f^{abc}$'s)
for an
odd $k$ vanish, at least
up to fairly high orders in perturbative
computations, and it is definitely true up to 3 loops.
This is because, $X_k^{a_1\dots a_k}$
can be expressed with only invariant tensors
$d_A^{b_1\dots b_j}$'s
($f^{abc}$'s can be eliminated) up
to certain high orders \cite{vanRitbergen:1998pn}.
Since $d_A^{a_1\dots a_j}=0$ for an odd $j$, 
$d_R^{a_1\dots a_k}X_k^{a_1\dots a_k}$
should vanish.\footnote{
Furthermore, if we take a sum $C_i(D)+C_i(\bar{D})$,
where $C_i(\bar{D}) = (-1)^n
{\rm Tr}({T^{a_n}_R\cdots T^{a_1}_R})\,
X_n^{a_1\dots a_n}$, in the sum, 
purely gluonic contributions to
$d_R^{a_1\dots a_k}X_k^{a_1\dots a_k}$ 
for an odd $k$ vanish to all orders.
Note that,
due to the charge conjugation symmetry,
if there is a color graph 
$C_i({D})$,
there is also a  conjugate graph
$C_i(\bar{D})$,
unless the two graphs are the same.
}

\begin{figure}[t]\centering
\begin{tabular}{ccc}
\includegraphics[width=3.5cm]{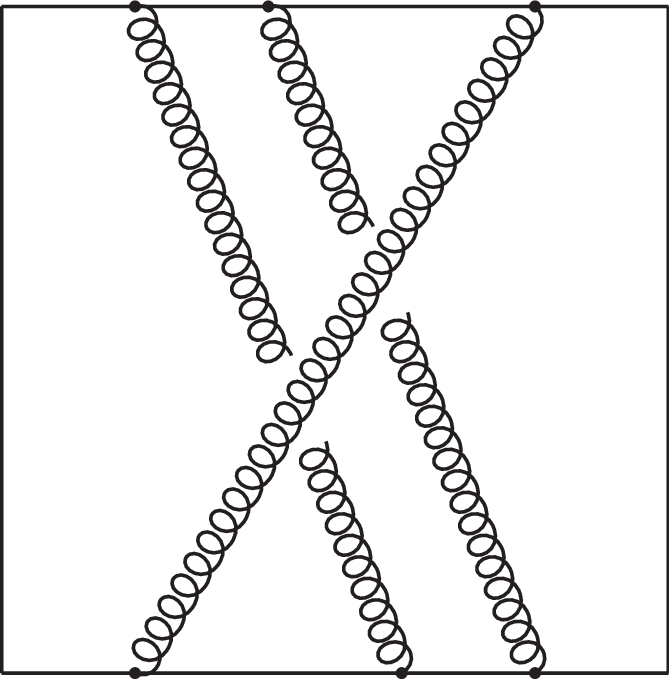}&&
\includegraphics[width=3.5cm]{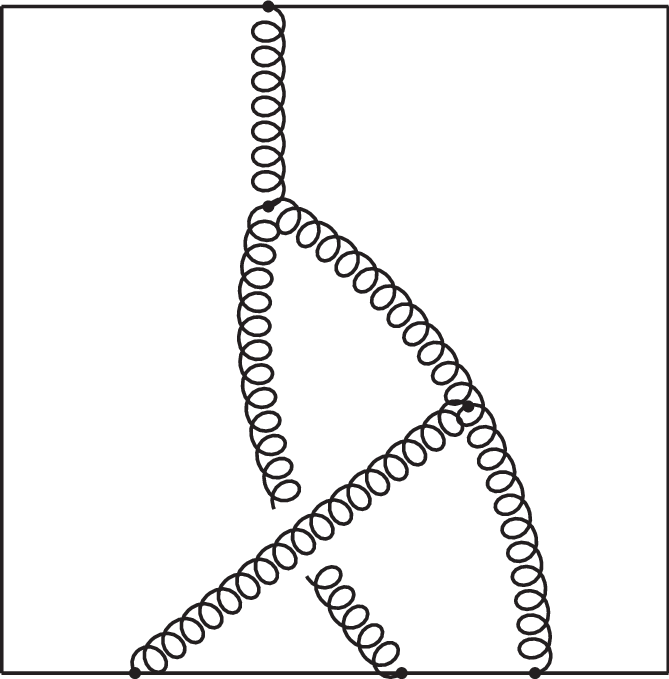}\\
(a)&~~~~~~~~~~~
&(b)
\end{tabular}
\caption{\small 
(a) 2-loop diagram $D$ with $n=6$, and (b) the corresponding color graph $C(D)$ 
with $n_c=4$.
Starting from the diagram (a), move the right-most vertex
on the upper line to the left by 
interchanging the vertices twice;
after removing the 2PR graphs, one is left with the
2PI color graph (b).
\label{2Loop-n=6}}
\end{figure}

Now we are in a position to discuss the loop diagrams.
We start from 2-loops.
Let us denote by $n$
the number of vertices on the Wilson loop in a 
2PI diagram $D$.
Then
we examine, for each fixed $n(\leq 6)$, the corresponding
color graphs.
Consider a diagram $D$ with $n=6$ or $n=5$. 
The diagram does not include fermion loops.
The number of 
vertices on the Wilson loop in 
the corresponding color graphs $C(D)$ is 
$n_c=4$; see Fig.~\ref{2Loop-n=6}.
The color graphs have a form
$C(D)={\rm Tr}({T^{a_1}_R\cdots T^{a_4}_R})\,
X_4^{a_1\dots a_4}$.
Each term of $X_4^{a_1\dots a_4}$ 
is expressed by two structure constants $f^{abc}f^{cde}$
with a different assignment of indices.
We rewrite ${\rm Tr}({T^{a_1}_R\cdots T^{a_4}_R})$
in terms of $d_R$'s using eq.~(\ref{prep1}).
Then $C(D)$ is expressed in the form of eq.~(\ref{geneformofC}).
There is, however, no way of contracting indices
such that $d_R^{abcd}$ survives.
Since $X_3^{abc}$ is
zero in the pure gluonic case, 
the only remaining term is $X_2^{aa}$.

We consider the diagrams $D$ with $n=4$.
The corresponding color graphs have either of the
following 3 forms: (i) 
$C(D)={\rm Tr}({T^{a_1}_R\cdots T^{a_4}_R})\,
X_4^{a_1\dots a_4}$ with
each term of $X_4^{a_1\dots a_4}$ 
expressed by two structure constants;
(ii) $C(D)={\rm Tr}(T^{a}_RT^{b}_RT^{c}_R)\,
X_3^{abc}$;
(iii) $C(D)={\rm Tr}(T^{a}_RT^{b}_R)\,
X_2^{ab}$.
The case (i) is the same as $n=5$,6.
In the case (ii) the purely gluonic contribution 
to the $d_R^{abc}X_3^{abc}$ term 
vanishes,
as we already noted;
on the other hand, the contributions of
graphs with
a fermion loop to $d_R^{abc}X_3^{abc}$ is proportional to
$f^{abc}d_R^{abc}=0$, since the fermion loop is included
as a gluon vacuum polarization.
The case (iii) is of the type
$X_2^{aa}$ by itself.
Thus, $C(D)$ has only the
$X_2^{aa}$ term in all three cases.

\begin{figure}[t]\centering
\includegraphics[width=15cm]{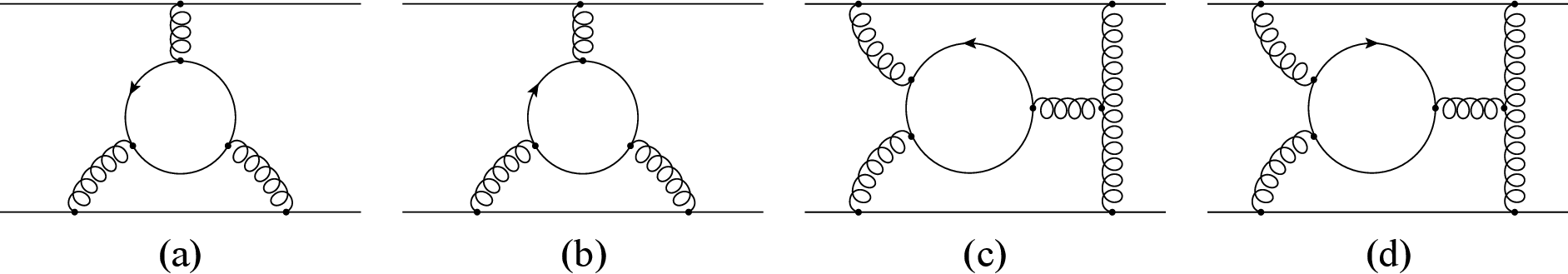}
\caption{\small 
Pairs of
color graphs with a fermion loop:
(a)(b) 2-loop graphs; (c)(d) 3-loop graphs.
The graphs in each pair have
the same topology and the opposite
charge flow directions in the fermion loops.
The $d_F^{abc}$ part of ${\rm Tr}(T_F^aT_F^bT_F^c)$
cancel between each pair.
\label{fermion-loop}}
\end{figure}

The color graphs, which corresponds to the diagrams with 
$n=3$, are either in the form
${\rm Tr}(T^{a}_RT^{b}_RT^{c}_R)\,X_3^{abc}$
or 
${\rm Tr}(T^{a}_RT^{b}_R)\,X_2^{ab}$.
The only new aspect, as compared to the cases
(ii) and (iii) of 
$n=4$, is in the diagrams including a fermion loop
which has three vertices on it.
These diagrams are proportional to
$n_l{\rm Tr}(T_F^aT_F^bT_F^c)=n_l(d_F^{abc}+\frac{i}{2}T_Ff^{abc})$.
The $n_ld_F^{abc}$ part is canceled between a pair of color graphs
with the same topology but with the opposite charge flows of
the fermion loop; see Figs.~\ref{fermion-loop}(a)(b).
This is because
${\rm Tr}(T_F^aT_F^bT_F^c)-{\rm Tr}(T_F^cT_F^bT_F^a)
= iT_Ff^{abc}$.
Thus, the sum of such a pair of color graphs is $n_lT_F$ times
the corresponding 1-loop color graph, in which the fermion
loop is replaced by the gluon three-point vertex.

The contributions of the diagrams with $n=2$ is of the
type $X_2^{aa}$.
Hence, all the contributions are reduced to 
$X_2^{aa}$ of 
eq.~(\ref{geneformofC}), which
is independent of $R$ apart from the overall factor $N_RC_R$.
Thus, the Casimir scaling holds.

We can discuss the 3-loop case in a similar manner.
As before, we discuss the color factors of 2PI diagrams
for each fixed $n$,
the number of vertices on the Wilson loop.
The maximum number is $n=8$.

For a moment, we consider diagrams without internal
fermion loops.
In the cases $n=8,7,6$ and 5,
the number of vertices on the Wilson loop in 
the corresponding color graphs $C(D)$ is
$n_c=5$ or less.
In the case $n_c=5$, 
the color graphs have a form
$C(D)={\rm Tr}({T^{a_1}_R\cdots T^{a_5}_R})\,
X_5^{a_1\dots a_5}$, and
each term of $X_5^{a_1\dots a_5}$ contains
$f^{a_ia_jb}$ ($\{a_i,a_j\}\subset\{a_1,\dots,a_n\}$).
According to eq.~(\ref{prep2}), the highest
rank invariant tensor in $C(D)$ is the 4th-rank
tensor $d_R^{abcd}$.
$C(D)$ includes the $X_2^{aa}$ and $X_4^{abcd}$ terms of
eq.~(\ref{geneformofC}).
Each term of $X_4^{abcd}$ includes four $f^{egh}=i(T_A^e)_{gh}$
and can be expressed as\footnote{
Many contractions are trivially zero, 
such as $f^{abe}d_R^{abcd}=0$, which are
not discussed explicitly.
} 
${\rm Tr}(T_A^{a'}T_A^{b'}T_A^{c'}T_A^{d'})$
with a different assignment of $(a',b',c',d')$ to $(a,b,c,d)$.
There are no contractions of the indices of $f^{egh}$'s
in the form
${\rm Tr}(T_A^{a'}T_A^{b'}){\rm Tr}(T_A^{c'}T_A^{d'})
\propto \delta^{a'b'}\delta^{c'd'}$;
these correspond to 2PR color graphs 
and would have generated $C_R^2$ had they contributed.
Using eq.~(\ref{prep1}) for $R=A$, 
$d_R^{abcd}X_4^{abcd}$ can be expressed by $d_R^{abcd}d_A^{abcd}$.
Thus, $C(D)$ is expressed by
$X_2^{aa}\propto C_A^3N_A$
and $X_4^{abcd}\propto d_A^{abcd}$.

The arguments in the other cases ($n=2,3,4$ and
$n_c\leq 4$ for $n=5,6,7,8$)
are the same as above, apart from the point
that for $n=2,3$ there is no $X_4^{abcd}$ term.
In the end the color factors are expressed
by $X_2^{aa}\propto C_A^3N_A$
and $X_4^{abcd}\propto d_A^{abcd}$.

Now we turn to the diagrams with fermion loops.
We denote
the number of vertices on a fermion loop by $n'(\leq 6)$.
A fermion loop with $n'=2$ is proportional to
$n_l{\rm Tr}(T_F^aT_F^b)=n_lT_F\delta^{ab}$.
Hence, if a 3-loop color graph includes at least 
one fermion loop with $n'=2$, it is equal to
$n_lT_F$ times the corresponding
2-loop graph,  in which the fermion
loop insertion is removed.
In this case, the $R$ dependence is only in the overall 
factor $N_RC_R$.

A fermion loop with $n'=3$ is proportional to
$n_l{\rm Tr}(T_F^aT_F^bT_F^c)$.
As in the 2-loop case,
the $n_ld_F^{abc}$ part is canceled between a pair of color graphs
with the same topology but with the opposite charge flows of
the fermion loop; see Figs.~\ref{fermion-loop}(c)(d).
Thus, the sum of such a pair of color graphs is $n_lT_F$ times
the corresponding 2-loop color graph, in which the fermion
loop is replaced by the gluon three-point vertex.

A fermion loop with $n'=4$ is proportional to
$n_l{\rm Tr}(T_F^aT_F^bT_F^cT_F^d)$,
$n_l{\rm Tr}(T_F^aT_F^bT_F^cT_F^b)$ or
$n_l{\rm Tr}(T_F^aT_F^bT_F^cT_F^c)$, and
a 3-loop color graph includes one such loop at most.
The latter two traces are proportional to $n_lT_F\delta^{ab}$
and represent gluon vacuum 
polarizations; namely the color factors in these cases
are proportional to those of
lower-loop graphs.
The first trace generates $n_ld_F^{abcd}d_R^{abcd}$ and lower rank tensors
via eq.~(\ref{prep1}).
Also in this case, the terms with $n_ld_F^{abc}$ are canceled between 
a pair of color graphs
with the same topology but with the opposite charge flows of
the fermion loop.
This  can be seen as follows.
Due to the hermiticity of $T_F^a$,
${\rm Tr}(T_F^aT_F^bT_F^cT_F^d)+{\rm Tr}(T_F^dT_F^cT_F^bT_F^a)
= 2\,{\rm Re}[{\rm Tr}(T_F^aT_F^bT_F^cT_F^d)]$.
If we substitute eq.~(\ref{prep1}),
the terms with odd numbers of $f^{egh}$ are pure imaginary
and vanish, hence
only the even rank tensors
$d_F^{abcd}$ and $d_F^{ab}=T_F\delta^{ab}$ remain.

A fermion loop with $n'=5$ is proportional to
${\rm Tr}(T_F^{a_1}\cdots T_F^{a_5})$ and included in diagrams with
$n=2$ and 3.
In most of the corresponding color graphs, 
at least one pair from the indices $(a_1,\dots ,a_5)$
is contracted between the pair;
in this case, the trace can be reduced to 
${\rm Tr}(T_F^aT_F^bT_F^c)$,
and the $n_ld_F^{abc}$ part is canceled between 
a pair of color graphs with the opposite charge flows of
the fermion loop, as before.
The only other type of contraction is in the form
${\rm Tr}(T_F^aT_F^bT_F^cT_F^dT_F^e)d_R^{ab}f^{cde}$,
which can be reduced to $d_R^{aa}$, i.e.\ the $X_2^{aa}$ term.

A fermion loop with $n'=6$ is contained only as a
gluon vacuum polarization in the
diagrams with $n=2$.
Hence, they contribute only to the $X_2^{aa}$ term
of eq.~(\ref{geneformofC}).

\begin{figure}[t]\centering
\includegraphics[width=15cm]{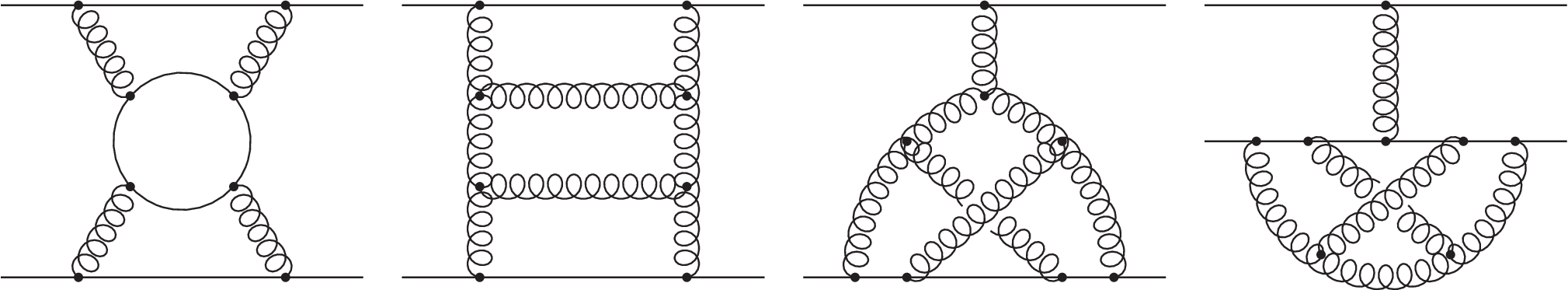}
\caption{\small 
Some of the
2PI diagrams which contribute to $d_R^{abcd}d_F^{abcd}$
and $d_R^{abcd}d_A^{abcd}$.
\label{diagrams-for-d44}}
\end{figure}

In summary, up to 3-loops
the Casimir scaling violating terms arise only as
$d_A^{abcd}d_R^{abcd}$ and $n_ld_F^{abcd}d_R^{abcd}$
in the $X_4^{abcd}$ term
of eq.~(\ref{geneformofC}).
Typical 2PI diagrams which contribute to these color factors
are shown in Fig.~\ref{diagrams-for-d44}.
All the other color factors are collected in the $X_2^{aa}$ term.
The absence of the odd-rank symmetric invariant tensors
$d_F,d_A$ follows from $(T^a_A)^T=-T^a_A$
and the cancellation between diagrams with opposite
charge flows in a fermion loop.
These features are ensured by the charge conjugation
symmetry.
Thus, the charge conjugation symmetry plays an important
role in suppressing the Casimir scaling violating
effects.


\section{\boldmath Numerical analysis of Casimir scaling violation}

In this section, we examine the perturbative predictions 
for the Casimir scaling violation
numerically and compare them
with lattice computations, in the case of QCD 
with zero flavor of quarks
(quenched approximation).
For the lattice results,
we will use those given
in \cite{Bali:2000un}, as they seem to be 
most accurate and extensive up to date.
Throughout the analysis,
we use the central value of
$r_0 \Lambda_{\overline{\rm MS}}^{\mbox{\scriptsize 3-loop}}
=0.574\pm 0.042
$ \cite{Sumino:2005cq}
to fix the relation between the lattice scale
 and
$\Lambda_{\overline{\rm MS}}^{\mbox{\scriptsize 3-loop}}$,
where $r_0$ denotes the Sommer scale \cite{Sommer:1993ce}.
(It is customary to interpret $r_0\approx 0.5$~fm
when comparing this scale to one of the real world.)
When making comparisons, one should note the
following point.
As is well known, the constant ($r$-independent)
part of the potentials by perturbative
and lattice computations cannot be related unambiguously.
This is because, the self-energy contribution and the potential
energy contribution in a lattice computation of the Wilson loop
cannot be separated unambiguously.
Furthermore, it is formidable in perturbation theory
to reproduce exactly 
the subtraction scheme used in \cite{Bali:2000un}, which subtracts
the leading-order self-energy contribution defined
in lattice perturbation
theory.
Alternatively, we study quantities which are,
by construction, independent
of the constant part of the potentials, and
discuss their relation to the 
Casimir-scaling violating effects measured in the lattice
simulation.

In this analysis we define the following two quantities as
measures for
violation of the Casimir scaling of the static potential
for the general representation $R$:
\begin{eqnarray}
\delta^{V_R}_{\rm CS}
&=&
\frac{[V_R(r)-V_R(r_1)]/C_R}{[V_F(r)-V_F(r_1)]/C_F}-1\,,
\\ 
\delta^{F_R}_{\rm CS}
&=&
\frac{V_R'(r)/C_R}{V_F'(r)/C_F}-1\,.
\end{eqnarray}
Both quantities are zero if the potential has an
exact Casimir
scaling property.
Both quantities are 
independent of the constant part of the
potential:
In the first quantity $\delta^{V_R}_{\rm CS}$,
we subtracted a constant from the potential $V_R(r)$
such that
it vanishes at $r=r_1$;
in the second quantity $\delta^{F_R}_{\rm CS}$, we used the
force $F_R(r) = -V_R'(r)$ between the color sources.
The fundamental representation ($R=F$) is used as a reference
in both quantities.

In principle $\delta^{V_R}_{\rm CS}$ and
$\delta^{F_R}_{\rm CS}$ can be measured in lattice
simulations.
On the other hand, they can be computed
in perturbation theory, by replacing
$V_{R,F}(r)$ by $[V_{R,F}(r)]_{\rm NNNLO}$.
They are equal at the leading order:
\begin{eqnarray}
\delta^{V_R}_{\rm CS}
,\delta^{F_R}_{\rm CS}
&=&
\left(\frac{\alpha_s(\mu)}{4\pi}\right)^3
\Biggl[\,
\frac{c_2n_l}{2N_A}\left(\frac{d_F^{abcd}d_R^{abcd}}{T_R}
-\frac{d_F^{abcd}d_F^{abcd}}{T_F} \right) 
\nonumber\\&&
~~~~~~~~~~~~~~~
+
\frac{c_4}{2N_A}\left(\frac{d_A^{abcd}d_R^{abcd}}{T_R}
-\frac{d_A^{abcd}d_F^{abcd}}{T_F} \right) 
\Biggr]
+{\cal O}(\alpha_s^4)\,.
\label{deltaR-pert}
\end{eqnarray}
At this order, $\delta^{V_R}_{\rm CS}$ and
$\delta^{F_R}_{\rm CS}$ are independent of $r$;
$\delta^{V_R}_{\rm CS}$ is also independent of $r_1$.

Eq.~(\ref{deltaR-pert}) shows that
$\delta^{V_R}_{\rm CS}$ and
$\delta^{F_R}_{\rm CS}$ are negative 
and their
absolute magnitudes increase with the color factor
$n_l\,d_F^{abcd}d_R^{abcd}/(N_AT_R)$ or with
$d_A^{abcd}d_R^{abcd}/(N_AT_R)$.
(Note that $c_2,c_4<0$.) 
As a result, the effect of Casimir scaling violation is to
reduce the tangent of $V_R(r)/C_R$ 
for larger $n_l\,d_F^{abcd}d_R^{abcd}/(N_AT_R)$ or for
larger $d_A^{abcd}d_R^{abcd}/(N_AT_R)$.
This can be seen from the fact that $\delta^{F_R}_{\rm CS}$, defined from
the static force, is reduced by this effect.

As shown in \cite{Anzai:2009tm},
in the case $G=SU(3)$, $R=F$ 
and $n_l=0$, the 
potential 
$[V_F(r)]_{\rm NNNLO}$ for the fundamental representation
agrees 
fairly well with the lattice results in the 
distance range $0.1\simlt r/r_0 \simlt 0.5$.\footnote{
It is estimated in  Ref.~\cite{Sumino:2005cq} 
that, by renormalization-group improvement 
 of $[V_R(r)]_{\rm NNNLO}$ (after subtraction
of the renormalon),
the agreement becomes better up to
larger distances, $r/r_0 \simlt 0.8$.
}
Hence, it would be natural to test whether the perturbative
prediction is also capable of reproducing the 
Casimir-scaling violating
effects in the same distance range.
Generally the perturbative predictions for $\delta^{V_R}_{\rm CS}$ and
$\delta^{F_R}_{\rm CS}$, eq.~(\ref{deltaR-pert}), 
are quite small, since they are 3-loop effects.
At the same time, they turn out to be
strongly scale-dependent, since only the leading-order terms 
with a high power of $\alpha_s$ are
known.
For example, for the adjoint (octet) potential of $SU(3)$
($R=A$),
we find
\bea
\delta^{V_A}_{\rm CS}
,\delta^{F_A}_{\rm CS}
&\approx& \alpha_s^3\,(-0.129-0.0030\,n_l)
\nonumber\\
&\approx&
\left\{
\begin{array}{lcl}
-0.00032&\rule[-5mm]{0mm}{6mm}~&
\left(
n_l=0,~
\mu=\mu_1=\Lambda_{\overline{\rm MS}}^{\mbox{\scriptsize 3-loop}}/0.035
\right)
\\
-0.0013&&
\left(
n_l=0,~
\mu=\mu_2=
\Lambda_{\overline{\rm MS}}^{\mbox{\scriptsize 3-loop}}/0.14
\right)
\end{array}
\right.
\,.
\label{est-adjoint-deltaCS}
\eea
Here, the scale $\mu_1$ ($\mu_2$) is the largest (smallest) scale,\footnote{
They 
correspond to $\mu_1^{-1}\approx 0.06\,r_0$ and
$\mu_2^{-1}\approx 0.24\,r_0$.
}
with which the stability of $[V_R(r)]_{\rm NNNLO}$ 
has been examined in \cite{Anzai:2009tm},
where $\mu_1$ and $\mu_2$ differ by a factor of 4.
It is a standard way to estimate uncertainties of 
the perturbative prediction for the potential.
Since $\delta^{V_A}_{\rm CS}$ and
$\delta^{F_A}_{\rm CS}$ are determined by the potential,
we use this scale dependence to estimate
uncertainties of these quantities.
Hence,
our estimates for $\delta^{V_A}_{\rm CS}$ and
$\delta^{F_A}_{\rm CS}$ are between $-0.03\%$ and
$-0.13\%$ for $n_l=0$,
in the 
distance range $0.1\simlt r/r_0 \simlt 0.5$. 
The large uncertainties in the estimates are due to the 
strong scale dependence.
The scale dependence is expected to be reduced
if the higher-order corrections (beyond 3-loop) to
$\delta^{V_A}_{\rm CS},\delta^{F_A}_{\rm CS}$ are
incorporated.

\begin{table}[t]
\begin{tabular}{ccccccc}
\hline\hline
~$N_R$~ &
$(p,q)~$&
$~C_R~~$ &
$~T_R~~$ &
$\frac{d_F^{abcd}d_R^{abcd}}{N_A T_R}$  &
$\frac{d_A^{abcd}d_R^{abcd}}{N_A T_R}$ &
$\delta^{V_R}_{\rm CS},\delta^{F_R}_{\rm CS}$(\%) for $n_l=0$ 
\\ \hline
$3$  &  $(1,0)$ & 4/3 & 1/2 & 5/48    & 15/8 &   \\
$8$  &  $(1,1)$ & 3   & 3   & 5/16    & 45/8 & $-0.03$ -- $-0.13$ \\
$6$  &  $(2,0)$ & 10/3& 5/2 & 17/48   & 51/8 &  $-0.04$ -- $-0.16$ \\
$~15_a$& $(2,1)$ & 16/3& 10& 29/48   & 87/8 & $-0.08$ -- $-0.31$ \\
$10$ &  $(3,0)$ &  6  & 15/2& 11/16   & 99/8 & $-0.09$ -- $-0.37$  \\
$27$ &  $(2,2)$ &  8  & 27& 15/16   & 135/8 & $-0.13$ -- $-0.52$ \\
$24$ &  $(3,1)$ &  25/3  & 25& 47/48   & 141/8 & $-0.13$ -- $-0.55$ \\
$~15_s$& $(4,0)$ & 28/3& 35/2& 53/48   & 159/8 & ~$-0.15$ -- $-0.63$
\vspace*{1.5mm}
\\
\vspace*{1.5mm}
$\Bigl[\frac{(k+1)(k+2)}{2}\Bigr]_s$ & $(k,0)$ & $\frac{k(k+3)}{3}$ &
$\frac{k(k+1)(k+2)(k+3)}{48}$ & $\frac{2k^2+6k-3}{48}$&
$\frac{3(2k^2+6k-3)}{8}$ &  
\hbox{\raise-4pt\hbox{$\stackrel{\scriptstyle (-0.0063 \mbox{ -- } -0.026)}
{\scriptstyle \times (k+4)(k-1)}$}}\\
\hline\hline
\end{tabular}
\caption{\small
Color factors 
for $SU(3)$, for the representations 
$R=3,8,6,15_a,10,27,24,15_s,\Bigl[\frac{(k+1)(k+2)}{2}\Bigr]_s$,
where each $R$ is labeled by its dimension
$N_R$. 
$(p,q)$ denotes the weight factor.
See Sec.~2 for the definitions of the other color
factors. 
The last column represents the estimates (in per cent) of 
$\delta^{V_R}_{\rm CS},\delta^{F_R}_{\rm CS}$ for $n_l=0$.
The last row represents the $k$th-rank completely symmetric
representation.
}
\label{tab:color2}
\end{table}

We list the color factors for various representations of
$SU(3)$ in Tab.~\ref{tab:color2}.
Our estimates for $\delta^{V_R}_{\rm CS}$,
$\delta^{F_R}_{\rm CS}$ for the corresponding representations
are shown in the same table for $n_l=0$.
All the estimates are derived in the same way
as in eq.~(\ref{est-adjoint-deltaCS}).
Note that in the case of $SU(3)$ simple relations
\bea
&&
\left[\frac{d_F^{abcd}d_R^{abcd}}{N_AT_R}\right]_{SU(3)}
\! \! =\frac{1}{8}
\left( \Bigl[C_R\Bigr]_{SU(3)}-\frac{1}{2} \right) ,
\label{SU3rel1}
\\ &&
\left[\frac{d_A^{abcd}d_R^{abcd}}{N_AT_R}\right]_{SU(3)}
\! \! =\frac{9}{4}
\left( \Bigl[C_R\Bigr]_{SU(3)}-\frac{1}{2} \right)
\label{SU3rel2}
\eea
hold, since $[d_F^{abcd}]_{SU(3)}=[d_A^{abcd}]_{SU(3)}/18=(\delta^{ab}\delta^{cd}+
\delta^{ac}\delta^{bd}+\delta^{ad}\delta^{bc})/24$.
It follows that the 
contribution of the internal quark per each flavor
is about 40 times smaller than that of the pure gluons,
independently of $R$.
Hence, the contributions of the internal quarks may be neglected
in QCD in the first approximation.
Eqs.~(\ref{SU3rel1}) and (\ref{SU3rel2}) also show that the Casimir-scaling
violations, $\delta^{V_R}_{\rm CS}$ and
$\delta^{F_R}_{\rm CS}$, scale proportionally to $C_R$
for $C_R \gg 1$ in QCD.

Let us compare our estimates with the lattice results.
We may relate $\delta^{V_R}_{\rm CS}$
to the Casimir scaling violating effects measured
in the lattice simulation.
From Fig.~4 and
the numbers listed in Tables II and IX of \cite{Bali:2000un}, 
$R_D(r)$ defined
in eq.~(31) of that paper is estimated to be close to
$(C_R/C_F)(1+\delta^{V_R}_{\rm CS})$ with a choice $r_1\sim 0.05\, r_0$.
Or equivalently,
\bea
\delta^{V_R}_{\rm CS}\approx\frac{C_F}{C_R}R_D-1 .
\eea
Since our prediction for $\delta^{V_R}_{\rm CS}$ is
independent of $r_1$ at leading order, we can compare
our predictions with the lattice results without
a precise knowledge of $r_1$.
Ref.~\cite{Bali:2000un} observed no violations of the Casimir scaling in
$R_D$ at $0.3\simlt r/r_0 \simlt 2$
within 5\% accuracy, for the representations
$R=8,6,15_a,10,27,24,15_s$.
This translates to 
bounds $|\delta^{V_R}_{\rm CS}|<5\%$ for all of these
representations in the same
distance range.
As seen in Tab.~\ref{tab:color2}, these bounds are perfectly consistent
with our predictions
in the overlapping distance range $0.3\simlt r/r_0 \simlt 0.5$:
our estimates are an order of magnitude smaller in the
largest cases.

It is desirable to test our predictions in
lattice simulations.
On the one hand, our current estimates for $\delta^{V_R}_{\rm CS}$,
$\delta^{F_R}_{\rm CS}$
have large uncertainties due to scale dependences,
and only their orders of magnitude can be predicted.
On the other hand,
the prediction, that these quantities scale proportionally
to specific combinations of color factors 
as given in eq.~(\ref{deltaR-pert}),
is expected to be more secure, especially for smaller $r$.
As already stated, the Casimir scaling violation is enhanced
for larger $C_R$ in QCD.
Hence, if a representation with an extremely large 
dimension is chosen, the violation may be detectable.
For instance, if we use the $k$th-rank completely symmetric
representation with $k=30$, the estimates for 
$\delta^{V_R}_{\rm CS}$,
$\delta^{F_R}_{\rm CS}$ amount to between $-6$\% and $-26$\%;
see Tab.~\ref{tab:color2}.

Another possible direction is to consider larger groups.
As can be seen in Tab.~\ref{tab:color},
the color factor $d_A^{abcd}d_A^{abcd}/(N_AT_A)$
increases rapidly with
$N$ both for $G=SU(N)$ and $G=SO(N)$.
However, the terms with
the leading power in $N$ cancel between
$d_A^{abcd}d_A^{abcd}/(N_AT_A)$ and $d_A^{abcd}d_F^{abcd}/(N_AT_F)$.
As a result $\delta^{V_A}_{\rm CS}$,
$\delta^{F_A}_{\rm CS}$ scale only linearly on $N$
for $SU(N)$ and quadratically for $SO(N)$.
Thus, even if a relatively large $N$ is chosen, it seems 
necessary to use higher rank representations in order
to detect Casimir scaling violating effect.

There is a way to perform an indirect test, presumably with
less efforts.
Although the leading $N$ dependence for $G=SU(N)$ or $G=SO(N)$ 
is canceled in 
$\delta^{V_A}_{\rm CS}$,
$\delta^{F_A}_{\rm CS}$,
this dependence is included in the fundamental potential $V_F(r)$.
In principle, it is possible to test the contribution 
of the $d_A^{abcd}d_F^{abcd}/(N_AT_F)$ term
in $V_F(r)$ by comparing lattice computations and
$[V_F(r)]_{\rm NNNLO}$.
In the case $G=SU(N)$, the contribution proportional
to $d_A^{abcd}d_F^{abcd}/(N_AT_F)$ 
in 
$[V_F(r)]_{\rm NNNLO}$ (or in $[dV_F(r)/dr]_{\rm NNNLO}$)
is estimated to be 
between $-6\%$ and $-23\%$ of the total amount for $N=25$, and 
between $-10\%$ and $-39\%$ for $N=30$,
using the same values for $\alpha_s(\mu_1)$ and $\alpha_s(\mu_2)$
as the $SU(3)$ case.
We may also use larger representations and test
the $d_A^{abcd}d_R^{abcd}/(N_AT_R)$  term.
Verifying the contribution of the $d_A^{abcd}d_R^{abcd}/(N_AT_R)$
term may be a first step toward
testing the Casimir scaling violation.

\section{Concluding remarks}

In this paper, we 
computed, within perturbation theory, 
the full ${\cal O}(\alpha_s^4)$ and
${\cal O}(\alpha_s^4\log\alpha_s)$ corrections to
the potential $V_R(r)$
between the static color sources, defined from the
Wilson loop
in a general representation $R$ of a general gauge group $G$.
The strictly perturbative contributions and
the ultra-soft contributions are separately presented.
The full expression for the
potential up to ${\cal O}(\alpha_s^4)$ 
and ${\cal O}(\alpha_s^4\log\alpha_s)$ is also given.

One way to utilize our result is as follows.
In the case $G=SU(3)$,
the present result provides many observables 
(corresponding to different $R$'s)
which can be used
in the matching procedure between
lattice QCD  and perturbative QCD.
Already there exist a number of computations of the 
potentials for various $R$'s
by lattice simulations.
Since the matching between
both theories using
the fundamental potential has been successful 
and can be used for the determination of $\alpha_s$
\cite{Sumino:2005cq}, 
it is expected
that our present result may 
be used in the same way and 
may contribute to improve
accuracies
in the determination of $\alpha_s$ or
in predicting other physical observables.

A prominent feature of our present result is that
the violation of
the Casimir scaling of the
potential is predicted at ${\cal O}(\alpha_s^4)$.
This is the first prediction,
based on a first-principle calculation, of the Casimir scaling
violation for the QCD potential.
The effect of the Casimir scaling violation is to
reduce the tangent of $V_R(r)/C_R$ 
proportionally to $d_A^{abcd}d_R^{abcd}/(N_AT_R)$ and to
$n_l\, d_F^{abcd}d_R^{abcd}/(N_AT_R)$.
We studied the sizes of the Casimir scaling violation
for various representations of the static sources
for $G=SU(3)$.
We find that they are fairly suppressed and fully
consistent with the present
bounds from lattice calculations, even 
if we take
into account strong scale-dependences of the predictions.
(Through a diagrammatic analysis, we observe that the
suppression of the Casimir scaling violation is
related to the charge conjugation symmetry.)

We also discussed possibilities that
our predictions for the effects
of the Casimir scaling violation
may be tested for representations with large
dimensions or in theories with larger gauge groups.
If these tests are unreachable in near future, one may 
alternatively perform
an indirect test of the contribution of the
$d_A^{abcd}d_R^{abcd}/(N_AT_R)$  term,
using the potentials
for various representations of large gauge groups.
If the perturbative prediction is excluded by 
lattice computations in these tests,
it means that (at least) for the Casimir scaling violating
effects non-perturbative contributions are 
dominant in the relevant
distance region.
Oppositely, suppose
the perturbative prediction is positively confirmed,
for instance, concerning the proportionality of the effects to the
specific color factors.
In this case,
the tiny Casimir scaling violation may serve as another
discriminant of models, which
attempt to explain the 
nature of QCD in the long-distance as well as 
in the intermediate-distance regions.

It is expected that the violation of the Casimir scaling of
the potential has some connection with the string breaking
phenomenon, since the Casimir scaling property
is known to be violated
in this phenomenon.
For this reason, it may be meaningful to focus on
the Casimir scaling violating effects in
the aforementioned class of models, even if
they are tiny at $r\simlt 1~{\rm fm}$.
In fact, the
string breaking phenomenon
is among the major dynamical issues of QCD between
the intermediate and long-distance regions to be
explained by these models.
We may entertain some speculation
from the perturbative computation.
The diagrams which contribute to 
the Casimir scaling violations include those
which (naively) are
relevant in the string breaking phenomenon
(for instance,
the first diagram of Fig.~\ref{diagrams-for-d44}
in the case $R=F$, and the second diagram in the
case $R=A$).
Moreover, the effect of the Casimir scaling violation is 
predicted to decrease the force between the static
charges, indicating that
screening of these charges is taking place. 
Although it is beyond the scope of perturbative QCD to 
treat the string breaking phenomenon,
some hints may be revealed by investigating further 
the Casimir scaling violation in the
distance region where perturbative treatment
is still valid.


\section*{Acknowledgements}
One of the authors (Y.S.) is grateful to A.~Hoang and 
T.~Onogi for fruitful discussion.
The work of Y.S.\ is supported in part by Grant-in-Aid for
scientific research No.\ 20540246 from
MEXT, Japan.

\end{document}